\documentclass[twocolumn,superscriptaddress,showpacs,amsmath,amssymb,pra,floatfix]{revtex4}

\usepackage{graphicx}
\usepackage{dcolumn}
\usepackage{bm}
\usepackage{bbm}
\usepackage{multirow,amssymb,amsbsy,amsmath}
\usepackage{stmaryrd}

\newcommand{\I}{\textup{i}}
\newcommand{\E}{\textup{e}}
\newcommand{\D}{\textup{d}}

\newcommand{\dd}{\text{d}}
\newcommand{\dod}[2]{\frac{\dd #1}{\dd #2}}

\newcommand{\ddim}{\udelta\kern0.1em}

\newcommand{\beikonst}[2]{\left( #1 \right)_{\kern-0.2em #2}}

\newcommand*{\ket}[1]{\mathopen{|}#1\mathclose{\rangle}}

\newcommand{\dop}{{\rho}}

\newcommand{\bk}{{\bf k}}
\newcommand{\bp}{{\bf p}}
\newcommand{\bq}{{\bf q}}
\newcommand{\dA}{\delta A}
\newcommand{\dAd}{\delta A^\dagger}
\newcommand{\dB}{\delta B}
\newcommand{\dBd}{\delta B^\dagger}

\begin{document}

\preprint{APS/123-QED}

%
%

\title{In situ measurement of the dynamic structure
  factor in ultracold quantum gases}

\author{Hendrik\ Weimer}%
\affiliation{Institute of Theoretical Physics III, Universit\"at Stuttgart, %
              70550 Stuttgart, Germany}%
\affiliation{Department of Physics, Harvard University, 17 Oxford Street, Cambridge, MA 02138, USA} %
\affiliation{ITAMP, Harvard-Smithsonian Center for Astrophysics, 60 Garden Street, Cambridge, MA 02138, USA}
\email{hweimer@cfa.harvard.edu}%
\author{Hans\ Peter\ B\"uchler}%
\affiliation{Institute of Theoretical Physics III, Universit\"at Stuttgart, %
              70550 Stuttgart, Germany}%

\date{\today}%

\begin{abstract}

  We propose an experimental setup to efficiently measure the dynamic
  structure factor of ultracold quantum gases. Our method uses the
  interaction of the trapped atomic system with two different cavity
  modes, which are driven by external laser fields. By measuring the
  output fields of the cavity the dynamic structure factor of
  the atomic system can be determined. Contrary to previous approaches
  the atomic system is not destroyed during the measurement process.

\end{abstract}


\pacs{67.85.De, 42.50.Ct, 42.50.Pq}
\maketitle

\section{Introduction}

The rapid experimental progress in the manipulation of ultracold
quantum gases has enabled the creation of strongly correlated
many-body systems, which are challenging to describe theoretically
\cite{Bloch2008}.  The unambiguous  identification of these phases
requires precise measurements to characterize its properties. 
 A powerful tool represents the response function, which
provides  static properties of the system as well as reveals information about  the excitations
of the examined state.  Here, we propose a method for an {\it in situ}
measurement of the response function to a external and weak 
probing field for a trapped gas of ultracold atoms.

In seminal experimental measurements the Bogoliubov excitation
spectrum in a Bose-Einstein condensate has been measured by probing
the dynamic structure factor \cite{Stamper-Kurn1999,Steinhauer2002}.
Nowadays, this method has been applied to access the excitation
spectrum within the BEC-BCS crossover \cite{Chin2004,stewart2008}, as
well as the smooth transition of the excitation gap from the
superfluid to the Mott insulating phase \cite{esslinger2004}.  A major
drawback of the present approaches is that the setup measures the
number of excitations created in the systems: the atomic system is
destroyed during the measurement process, and in addition, it requires
the application of strong fields violating in most situations the
condition of a weak probe.  As a consequence, it becomes difficult to
distinguish, whether the measurements probes the dynamical evolution
of a strongly driven system or characterizes the ground state
properties.  Several alternative methods to access information on
ground state properties by measuring the light field passing through
an atomic system have been discussed, e.g., the detection of magnetic
order \cite{eckert2007,eckert2008,vega2008,douglas2010} and particle
fluctuations \cite{mekhov2007,mekhov2009}.  Within a remarkable
experiment, the possibility to access the dynamical structure factor
by a measurement of the probe field has recently been demonstrated
\cite{Pino2011}, which opens a way for weakly probing and analyzing
cold atomic gases.

\begin{figure}[tb]
  \includegraphics[width=0.95\linewidth]{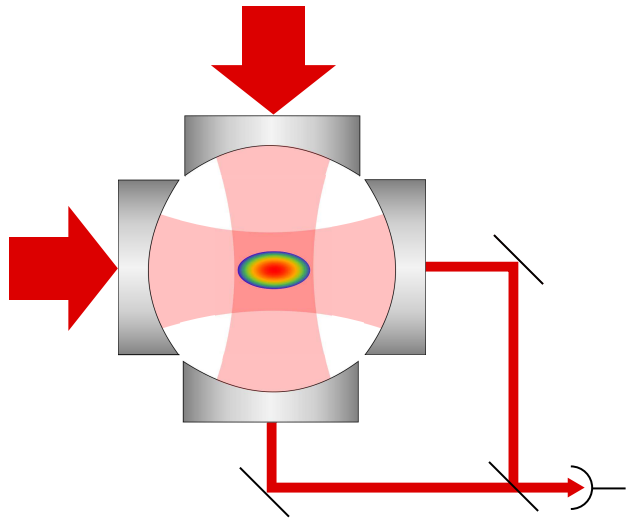}
  \caption{Proposed experimental setup: an ultracold atomic cloud
    interacts with two different cavity fields, which are driven by
    external lasers. The light fields leaving the cavity contain
    information on the dynamic structure factor of the atomic system,
    which can be recovered in a homodyne detection scheme.}
  \label{fig:setup}
\end{figure}

In this article, we propose a setup for realizing an {\it in-situ}
measurement of the linear response function for a cloud of trapped
ultracold atomic atoms.  The setup uses two different cavity fields,
see Fig~\ref{fig:setup}, and is motived by recent progress of
Bose-Einstein condensates coupled to high finesse cavities
\cite{Slama2007,Brennecke2007,Colombe2007,Murch2008}. Note that the
experiment can also be performed in a ring cavity setup to avoid
standing waves created by the probe lasers; similar setups have been
proposed for cavity state preparation \cite{mattinson01,larson05}. The
main idea is that the interaction of the atomic system with the cavity
modes transfers information on the response of the atomic system onto
the cavity fields \cite{mekhov2007,eckert2008}. By measuring the
fields that leave the cavities one can reconstruct the linear response
function as well as the dynamic structure factor.  Our method puts
only modest requirements on the cavity finesse and allows for fast
measurements with sufficiently high photon count rates without
destroying the atomic system.

For simplicity, we focus in the following on the density response function; the extension to
alternative setups accessing e.g., the  spin structure factor is straight forward.
For weak probe fields $\phi$, the linear response function is defined by the
deformation of the particle density
\begin{equation}
  \langle \rho_{\bf q}(\omega)\rangle = \chi_{\bf q}(\omega) \phi_{\bf q}(\omega) 
  \end{equation}
with the Fourier transformation 
relation $\phi_{\bf q}(\omega) = \int dt d{\bf r} \phi(t,{\bf r}) \exp(i \omega t-i {\bf q} {\bf r})$.
While the real part of the response function $\chi'$ accounts for the dispersive properties of the media
on the probe field, the imaginary part $\chi''$ describes the creation of excitations within the media.
Consequently, the response function contains important information about the 
two-particles excitations as well as collective excitations, which is obvious from
its relation to the dynamical structure factor
\begin{equation}
  \chi''_\bq(\omega) = -\pi[1-\exp(-\beta\hbar\omega)]S(\bq,\omega).
\end{equation}
In previous experiments the dynamic structure factor $S(\bq,\omega)$
has been studied extensively. However, it is important to point out,
that the real part and the imaginary part of the response function are
not independent of each other, but rather are related via the
Kramers-Kronig relation
\begin{equation}
  \chi''_q(\omega) = -\frac{2}{\pi}\omega\int\limits_0^\infty \D\omega'\frac{\chi'_q(\omega')}{\omega'^2-\omega^2}  \label{kramerskronig}.
\end{equation}
As a consequence, it is possible to access information about the excitation spectrum by probing
the real part of the response function off-resonantly and without creating any excitations in the media.
Such a method is in strong contrast to the current approach, where the created excitations strongly
distort the probed state. The method present in this manuscript allows one to access both the real
part of  the response function as well as the imaginary part, and opens a way to study {\it in situ}
and for very weak probes the response of an quantum many-body system with
cold atomic and molecular gases. 

\section{Hamiltonian description}

\begin{figure}[tb]
  \includegraphics[width=0.8\linewidth]{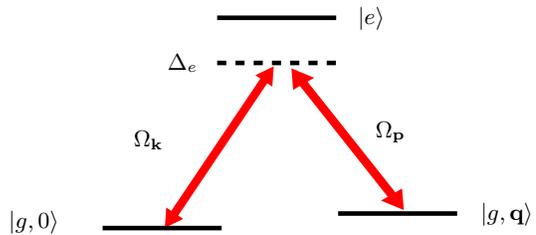}
  \caption{Internal level structure of the atomic system. Transitions
    between the ground state $\ket{g}$ and the excited state $\ket{e}$
    are driven by cavity fields $\Omega_\bk$, $\Omega_\bp$ with a
    large detuning $\Delta_e$ from the atomic resonance.  The momentum
    difference ${\bf q} = \bk - \bp$ of the cavity fields is
    transfered to the atomic system.}
  \label{fig:excite}
\end{figure}

We consider a system of $N$ atoms with two internal states
$\ket{g}$ and $\ket{e}$ (see Fig.~\ref{fig:excite}), coupled to two
cavity fields  $a_{\bf k}$ and $a_{\bf p}$ with frequencies 
$\omega_{\bf k}$ and $\omega_{\bf p}$, respectively. 
The dynamics of the cavity modes is governed 
by the Hamiltonian
\begin{equation}
  H_C = \hbar\omega_\bk a_\bk^\dagger a_\bk + \hbar\omega_\bp a_\bp^\dagger a_\bp.
\end{equation}
For a large detuning $\Delta_e$ of the cavity fields from the atomic
resonance, we can adiabatically eliminate the state
$\ket{e}$. Ignoring effects from spontaneous emission, the interaction
between the two cavity fields and the particle density operator $\rho_{\bf q}$ of 
the atomic system takes the form
\begin{equation}
  H_I = g\left[\rho_0\left(a_\bk^\dagger a_\bk+ a_\bp^\dagger a_\bp\right) + \rho_{-\bf q}a_\bk^\dagger a_\bp + \rho_{\bf q}a_\bp^\dagger a_\bk\right],
\end{equation}
with $g$ being the two-photon coupling constant, while ${\bf q} = \bk -
\bp$ denotes the difference in momenta between the cavity fields. The
first term in $H_I$ accounts for the Stark shift of the atoms and only
leads to a renormalization of the cavity resonance frequencies, i.e.,
$\bar{\omega}_{\bk,\bp} = {\omega}_{\bk,\bp}+\dop_0 g/\hbar$. Note
that here we assume $\dop_0$ denotes the total number of atoms in the
cavity and is a conserved quantity. In addition we would also like to point out that the cavity lasers can also be focused and
probe only parts of the atomic system. Then, $\rho_{0}$ accounts only for the
atoms within the mode volume of the lasers, and the Stark shift becomes a local
shift in the chemical potential.

\section{Weak coupling expansion}

\subsection{Free field solution}

We consider the terms involving the density operator $\dop_{\pm\bf q}$
as small perturbations and solve the system first in absence of these
coupling terms. Then, the quantum Langevin equation for the cavity
fields reduce to
\begin{eqnarray}
  \dot{a}_\bk &=& -\I\bar{\omega}_\bk a_\bk -
  \frac{\gamma}{2}a_\bk+\sqrt{\gamma}b_\bk^{\mathrm{in}}(t)\\
  \dot{a}_\bp &=& -\I\bar{\omega}_\bp a_\bp -
  \frac{\gamma}{2}a_\bp+\sqrt{\gamma}b_\bp^{\mathrm{in}}(t),
\end{eqnarray}
where $b_{\bk,\bp}^{\mathrm{in}}(t)$ are the input fields of the
cavities \cite{Walls1994}. We assume that the cavity is driven by two
lasers with frequencies $\Omega_\bk$ and $\Omega_\bp$, i.e., $\langle
b^{\mathrm{in}}_{\bk,\bp}(t)\rangle =
\beta_{\bk,\bp}\exp(-\I\Omega_{\bk,\bp}t)$. Then, the steady state
solution for the averaged fields $\alpha_{\bk,\bp} = \langle
a_{\bk,\bp}\rangle$ is given by
\begin{eqnarray}
  \alpha_\bk(t) &=& \frac{\sqrt{\gamma}}{\gamma/2+\I(\bar{\omega}_\bk-\Omega_\bk)}\beta_\bk\exp(-\I\Omega_\bk t)\\
  \alpha_\bp(t) &=& \frac{\sqrt{\gamma}}{\gamma/2+\I(\bar{\omega}_\bp-\Omega_\bp)}\beta_\bp\exp(-\I\Omega_\bp t).
\end{eqnarray}
To simplify the analysis, we set the cavity detunings to zero, $\Delta =
\bar{\omega}_\bk-\Omega_\bk=\bar{\omega}_\bp-\Omega_p=0$ and assume
equal driving fields $|\beta_\bk|=|\beta_\bp|$, while the
generalization to unequal driving fields is
straightforward. Furthermore, we can absorb a phase factor in the
definition of $\beta_{\bk,\bp}$ such that $\alpha_\bk(t) =
\alpha_0\exp(-\I\Omega_\bk t)$ and $\alpha_\bp(t) =
\alpha_0\exp(-\I\Omega_\bp t)$ with $\alpha_0 =
2|\beta_\bk|/\sqrt{\gamma}$.

\subsection{Effects of fluctuations}

To study the effect of the atom-field interaction, we consider small
fluctuations around the free field solution. For this, we define a new
set of operators as
\begin{eqnarray}
  A &=& \frac{1}{\sqrt{2}}\left(a_{\bk}\E^{\I\Omega_\bk t}+a_{\bp}\E^{\I\Omega_\bp t}\right)\\
  B &=& \frac{1}{\sqrt{2}}\left(a_{\bk}\E^{\I\Omega_\bk t}-a_{\bp}\E^{\I\Omega_\bp t}\right)
\end{eqnarray}
with the inverse transformation
\begin{eqnarray}
  a_\bk = \frac{1}{\sqrt{2}}(A+B)\E^{-\I\Omega_\bk t}\\
  a_\bp = \frac{1}{\sqrt{2}}(A-B)\E^{-\I\Omega_\bp t}.
\end{eqnarray}
These operators have the expectation values $\langle A\rangle =
2\sqrt{2/\gamma}\beta_\bk$ and $\langle B \rangle = 0$, and we expand them
into their mean values and small fluctuations $\dA$, $\dB$. Then, the
interaction Hamiltonian $H_I=H_1+H_2+H_3$ reduces to
\begin{eqnarray}
  H_1 &=& \int \D{\bf x} \rho({\bf x}) V_0 \cos({\bf qx}-\omega t)\\
  H_2 &=& \frac{g_{\mathrm{eff}}}{2}\left(\rho_\bq\E^{-\I\omega t}+\rho_{-\bq}\E^{\I\omega t}\right)\left(\dA+\dAd\right)\\
  H_3 &=& \frac{g_{\mathrm{eff}}}{2}\left(\rho_\bq\E^{-\I\omega t}-\rho_{-\bq}\E^{\I\omega t}\right)\left(\dB-\dBd\right)
\end{eqnarray}
with $\omega = \Omega_\bk-\Omega_\bp$, the potential strength $V_0 =
g|\langle A\rangle|^2$, and the effective coupling $g_{\mathrm{eff}} =
g|\langle A\rangle|$. The first term describes a classical driving
field $V_{\mathrm{ext}} = V_0\cos({\bf qx}-\omega t)$ for the atomic
system, while the last terms account for the coupling between the
atomic system and the cavity fields.  For $\omega = 0$, this set of
equations reduces to the case previously studied in the context of
cavity cooling \cite{Griessner2004}. We now introduce the quadrature
operators
\begin{align}
  X_B &= \frac{1}{2}\left[\dB + \dBd\right]\hspace{0.5cm} & P_B &= -\I\left[\dB-\dBd\right]\\
 X_A &= \frac{1}{2}\left[\dA + \dAd\right] & P_A &= -\I\left[\dA-\dAd\right]
\end{align}
with the commutation relations
\begin{eqnarray}
  [X_A,P_A] &= [X_B,P_B] =& \I\\{}
  [X_B,P_A] &= [X_A,P_B] =& 0.
\end{eqnarray}
The equation of motions for the cavity field then reduce to
\begin{eqnarray}
  \dot{X}_A &=& -\frac{\gamma}{2}X_A \\
  \dot{P}_A &=& -\frac{g_{\mathrm{eff}}}{\hbar}[\rho_\bq\E^{-\I\omega t}+\rho_{-\bq}\E^{\I\omega t}]-\frac{\gamma}{2}P_A\\
  \dot{X}_B &=& \I\frac{g_{\mathrm{eff}}}{\hbar}[\rho_\bq\E^{-\I\omega t}-\rho_{-\bq}\E^{\I\omega t}]-\frac{\gamma}{2}X_B\\
  \dot{P}_B &=& -\frac{\gamma}{2}P_B.
\end{eqnarray}
These equations describe the back action of the atomic system with the particle density $\rho_{\bf q}$ onto the cavity fields, i.e.,
a fluctuation in the particle density operator of the atomic system $\rho_{\bf q}(t)$ influences the 
cavity fields within characteristic quantum-non-demolition setup. 

\subsection{Linear response regime}

The leading term $H_{1}$ drives a perturbation onto density of the
atomic system with the field $\phi(t,{\bf x}) = V_{0} \cos({\bf q}
{\bf x} - \omega t)$. The condition of a weak probe field reduces to
$V_{0}< E_{s}$ with $E_{s}\sim \hbar^2 {\bf q}^2/m$ the characteristic
energy scale of the atomic system \cite{VanHove1954}. Then, the
response of the atomic system to this external probe is well described
within linear response theory
\begin{equation}
  \langle  \rho(t,{\bf x})\rangle  = \int dt' d{\bf x}'  \chi(t-t',{\bf x}-{\bf x}') \phi(t',{\bf x'}),
\end{equation}
which implies for the present drive with momentum transfer ${\bf q}$ and 
frequency $\omega$ the response
$\langle \rho_{\bf q}(\omega )\rangle = V_{0} \chi_{\bf q}(\omega)/2 $.
Then, it is possible to replace the
operators in the Langevin equations by their expectation values, and
we obtain the coupled differential equations
\begin{eqnarray}
  \dod{}{t}\langle{X}_A\rangle &=& -\frac{\gamma}{2}\langle X_A\rangle \\
  \dod{}{t}\langle{P}_A\rangle &=& -\frac{g_{\mathrm{eff}}}{\hbar}V_0\chi'_\bq(\omega)-\frac{\gamma}{2}\langle P_A\rangle\\
  \dod{}{t}\langle{X}_B\rangle &=& -\frac{g_{\mathrm{eff}}}{2\hbar}V_0\chi''_\bq(\omega)-\frac{\gamma}{2}\langle X_B\rangle\\
  \dod{}{t}\langle{P}_B\rangle &=& -\frac{\gamma}{2}\langle P_B\rangle.
\end{eqnarray}
We can immediately identify the steady state solution
\begin{eqnarray}
  \langle X_A\rangle &=& \langle P_B\rangle = 0\\
  \langle{P}_A\rangle &=& -\frac{2g_{\mathrm{eff}}V_0}{\hbar\gamma} \chi'_\bq(\omega)\\
\langle{X}_B\rangle &=& -\frac{g_{\mathrm{eff}}V_0}{\hbar\gamma} \chi''_\bq(\omega).
\end{eqnarray}
Consequently, the linear response of the atomic system is imprinted
onto the cavity fields: the imaginary part of the response function describing
the creation of two-particle excitations and collective excitations in the atomic system
is encoded by a shift in the amplitude operator $X_{B}$ accounting for the scattered of 
photons from one cavity mode onto the other. In turn, the real part of the response function
characterizing the dispersive part of the media leads to a shift in the phase $P_{A}$ of the fields.

\section{Experimental parameters}

Finally, we provide the experimental parameters for the measurement of the
response function. We will focus on the measurement of  the phase quadrature
$\langle P_A \rangle$, as it exhibits  two advantages over
the measurement of $\langle X_B \rangle$: First, we note that $\langle
P_A \rangle$ is directly given by the sum of the expectation values
$\langle P_\bk \rangle$ and $\langle P_\bp \rangle$ of the physical
cavity fields $a_\bk$ and $a_\bp$, which can be efficiently
measured within a homodyne detection. For appropriate phases of the
cavity fields this means that the measurements are done against zero
background, which is not the case when measuring $\langle X_B
\rangle$. Second, the Kramers-Kronig relation Eq.~(\ref{kramerskronig})
allows us to compute $\chi''_q(\omega)$ {\em on resonance} by using
measurement data taken {\em off resonance}. Therefore, we are able to
obtain the dynamic structure factor of the atomic system without
transferring energy onto the atomic system.

\begin{figure}[b]
  \includegraphics[width=0.8\linewidth]{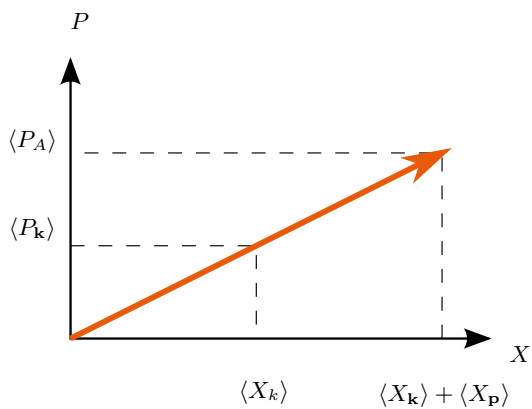}
  \caption{Phasor diagram illustrating the measurement of $\langle
    P_A\rangle$ as the sum of the physical observables $\langle
    P_\bk\rangle$, $\langle P_p\rangle$. A signal can be detected only
    when the ratio $\langle P_A\rangle/(\langle X_\bk\rangle + \langle
    X_\bp\rangle$ is large compared to phase fluctuations.}
  \label{fig:paxa}
\end{figure}

The validity of the expansion and ignoring the quadratic terms in the fluctuating fields
requires a large photon number in the cavity, i.e., $\langle A\rangle \gg 1$. In turn, the
weak driving field $V_{0} \lesssim E_{s}$ implies a weak coupling 
$g = V_{0}/|\langle A \rangle|^2 \lesssim E_{s}/|\langle A \rangle|^2$. Such weak coupling
is easily achieved by controlling the detuning to the excited level, see Fig.~\ref{fig:excite}.  Note, that his weak coupling is in high contrast
to the recent experimental works on high finess cavities \cite{Slama2007,Brennecke2007,Colombe2007,Murch2008}, and significantly simplifies  the requirements for the experimental realization of such a setup.
The phase quadratures
can be efficiently been measured within a homodyne detection with the sensitivity, see Fig.~\ref{fig:paxa},
\begin{equation}
   S =  \frac{\langle P_{A}\rangle}{\langle A\rangle}= \frac{\langle P_{\bf k}\rangle+ \langle P_{\bf p}\rangle}{\langle A\rangle}.
\end{equation}
The range of interesting values for the response function are 
$\chi_{\bf q}(\omega) \approx N /E_{s}$ with $N$ the total number of particles 
and $E_{s}$ the characteristic energy scale of the atomic system. This requires
a sensitivity to measure the phase angle
\begin{equation}
  S \approx \frac{N g}{\hbar \gamma}  \frac{V_{0}}{E_{s}} = \frac{V_{0}}{\hbar \gamma} \frac{N}{|\langle A\rangle|^2} \frac{V_{0}}{E_{s}}
\end{equation}
For the last equation, we have used  the relation between the driving field 
$V_{0} = g |\langle A\rangle|^2$ and the number of photons in the cavity.
The number of atoms within the cavity are typically in the range $N\sim 10^6$,
and therefore, with a cavity decay rate $\gamma \sim 100\,{\rm GHz}$ corresponding to a quality factor $Q\sim 10^4$
with up to 100 photons within the cavity, and a characteristic energy $E_{s} \sim 1\,{\rm KHz}$,
the required sensitivity is in the range $S\sim 10^{-4}$, which can be achieved with present techniques \cite{Hansen2001}.

\section{Conclusions}

In summary we have demonstrated the experimental feasibility of an {\it in
situ} measurement of the dynamic structure factor of arbitrary atomic
systems, providing a novel tool for the coherent manipulation of
ultracold quantum gases. The proposed method employs two cavity fields
which are driven by external lasers. We have shown that the required
resolution can be achieved using a cavity finesse that has already
been reached in present experiments. Furthermore, the method allows
for sufficiently fast measurements.

The current setup is in close analogy to the recently experimentally realized 
system \cite{Pino2011}. The main difference is, that in the experimental setup a 
strong imbalance between the two photon modes is applied. As a consequence,
the validity of the expansion up to linear order in the fluctuating fields is satisfied
by the strong coupling laser, while for the weak probe beam a very low photon 
number can be used. Then, the present calculations
can be performed again in a straightforward manner with the only modification, that
the real part of the response function is written on the phase quadrature $P_{\bf p}$, while
the imaginary part is encoded onto the amplitude quadrature $X_{\bf p}$.  In the experiment
the amplitude quadrature was accessed with a heterodyne detection \cite{Pino2011}, while the
real part of the response function again could be probed by a simpler homodyne detection.
We would like to point out that the contrast of the signal can be strongly enhanced by
using a cavity for the weak probe field.

Finally, we would like to point out, that the present setup can also be used to study the
relaxation of excitations in the atomic system: first, the lasers imprint several excitations
on resonance into the atomic system, while at a later stage the probe of the real part allows
one to analyze whether these excitations are still present or have relaxed. 

\begin{acknowledgments}

We thank R. L\"ow for fruitful discussions on the experimental
parameters. The work was supported by a grant from the Army Research
Office with funding from the DARPA OLE program, by the Deutsche
Forschungsgemeinschaft (DFG) within SFB/TRR 21, by the National
Science Foundation through a grant for the Institute for Theoretical
Atomic, Molecular and Optical Physics at Harvard University and
Smithsonian Astrophysical Observatory, and by a fellowship within the
Postdoc Program of the German Academic Exchange Service (DAAD).

\end{acknowledgments}


\begin{thebibliography}{20}
\expandafter\ifx\csname natexlab\endcsname\relax\def\natexlab#1{#1}\fi
\expandafter\ifx\csname bibnamefont\endcsname\relax
  \def\bibnamefont#1{#1}\fi
\expandafter\ifx\csname bibfnamefont\endcsname\relax
  \def\bibfnamefont#1{#1}\fi
\expandafter\ifx\csname citenamefont\endcsname\relax
  \def\citenamefont#1{#1}\fi
\expandafter\ifx\csname url\endcsname\relax
  \def\url#1{\texttt{#1}}\fi
\expandafter\ifx\csname urlprefix\endcsname\relax\def\urlprefix{URL }\fi
\providecommand{\bibinfo}[2]{#2}
\providecommand{\eprint}[2][]{\url{#2}}

\bibitem[{\citenamefont{Bloch et~al.}(2008)\citenamefont{Bloch, Dalibard, and
  Zwerger}}]{Bloch2008}
\bibinfo{author}{\bibfnamefont{I.}~\bibnamefont{Bloch}},
  \bibinfo{author}{\bibfnamefont{J.}~\bibnamefont{Dalibard}}, \bibnamefont{and}
  \bibinfo{author}{\bibfnamefont{W.}~\bibnamefont{Zwerger}},
  \bibinfo{journal}{Rev. Mod. Phys.} \textbf{\bibinfo{volume}{80}},
  \bibinfo{eid}{885} (\bibinfo{year}{2008}).

\bibitem[{\citenamefont{Stamper-Kurn et~al.}(1999)\citenamefont{Stamper-Kurn,
  Chikkatur, G\"orlitz, Inouye, Gupta, Pritchard, and
  Ketterle}}]{Stamper-Kurn1999}
\bibinfo{author}{\bibfnamefont{D.~M.} \bibnamefont{Stamper-Kurn}},
  \bibinfo{author}{\bibfnamefont{A.~P.} \bibnamefont{Chikkatur}},
  \bibinfo{author}{\bibfnamefont{A.}~\bibnamefont{G\"orlitz}},
  \bibinfo{author}{\bibfnamefont{S.}~\bibnamefont{Inouye}},
  \bibinfo{author}{\bibfnamefont{S.}~\bibnamefont{Gupta}},
  \bibinfo{author}{\bibfnamefont{D.~E.} \bibnamefont{Pritchard}},
  \bibnamefont{and} \bibinfo{author}{\bibfnamefont{W.}~\bibnamefont{Ketterle}},
  \bibinfo{journal}{Phys. Rev. Lett.} \textbf{\bibinfo{volume}{83}},
  \bibinfo{pages}{2876} (\bibinfo{year}{1999}).

\bibitem[{\citenamefont{Steinhauer et~al.}(2002)\citenamefont{Steinhauer,
  Ozeri, Katz, and Davidson}}]{Steinhauer2002}
\bibinfo{author}{\bibfnamefont{J.}~\bibnamefont{Steinhauer}},
  \bibinfo{author}{\bibfnamefont{R.}~\bibnamefont{Ozeri}},
  \bibinfo{author}{\bibfnamefont{N.}~\bibnamefont{Katz}}, \bibnamefont{and}
  \bibinfo{author}{\bibfnamefont{N.}~\bibnamefont{Davidson}},
  \bibinfo{journal}{Phys. Rev. Lett.} \textbf{\bibinfo{volume}{88}},
  \bibinfo{pages}{120407} (\bibinfo{year}{2002}).

\bibitem[{\citenamefont{Chin et~al.}(2004)\citenamefont{Chin, Bartenstein,
  Altmeyer, Riedl, Jochim, Denschlag, and Grimm}}]{Chin2004}
\bibinfo{author}{\bibfnamefont{C.}~\bibnamefont{Chin}},
  \bibinfo{author}{\bibfnamefont{M.}~\bibnamefont{Bartenstein}},
  \bibinfo{author}{\bibfnamefont{A.}~\bibnamefont{Altmeyer}},
  \bibinfo{author}{\bibfnamefont{S.}~\bibnamefont{Riedl}},
  \bibinfo{author}{\bibfnamefont{S.}~\bibnamefont{Jochim}},
  \bibinfo{author}{\bibfnamefont{J.~H.} \bibnamefont{Denschlag}},
  \bibnamefont{and} \bibinfo{author}{\bibfnamefont{R.}~\bibnamefont{Grimm}},
  \bibinfo{journal}{Science} \textbf{\bibinfo{volume}{305}},
  \bibinfo{pages}{1128} (\bibinfo{year}{2004}).

\bibitem[{\citenamefont{Stewart et~al.}(2008)\citenamefont{Stewart, Gaebler,
  and Jin}}]{stewart2008}
\bibinfo{author}{\bibfnamefont{J.~T.} \bibnamefont{Stewart}},
  \bibinfo{author}{\bibfnamefont{J.~P.} \bibnamefont{Gaebler}},
  \bibnamefont{and} \bibinfo{author}{\bibfnamefont{D.~S.} \bibnamefont{Jin}},
  \bibinfo{journal}{Nature} \textbf{\bibinfo{volume}{454}},
  \bibinfo{pages}{744} (\bibinfo{year}{2008}).

\bibitem[{\citenamefont{St\"oferle et~al.}(2004)\citenamefont{St\"oferle,
  Moritz, Schori, K\"ohl, and Esslinger}}]{esslinger2004}
\bibinfo{author}{\bibfnamefont{T.}~\bibnamefont{St\"oferle}},
  \bibinfo{author}{\bibfnamefont{H.}~\bibnamefont{Moritz}},
  \bibinfo{author}{\bibfnamefont{C.}~\bibnamefont{Schori}},
  \bibinfo{author}{\bibfnamefont{M.}~\bibnamefont{K\"ohl}}, \bibnamefont{and}
  \bibinfo{author}{\bibfnamefont{T.}~\bibnamefont{Esslinger}},
  \bibinfo{journal}{Phys. Rev. Lett.} \textbf{\bibinfo{volume}{92}},
  \bibinfo{pages}{130403} (\bibinfo{year}{2004}).

\bibitem[{\citenamefont{Eckert et~al.}(2007)\citenamefont{Eckert, Zawitkowski,
  Sanpera, Lewenstein, and Polzik}}]{eckert2007}
\bibinfo{author}{\bibfnamefont{K.}~\bibnamefont{Eckert}},
  \bibinfo{author}{\bibfnamefont{L.}~\bibnamefont{Zawitkowski}},
  \bibinfo{author}{\bibfnamefont{A.}~\bibnamefont{Sanpera}},
  \bibinfo{author}{\bibfnamefont{M.}~\bibnamefont{Lewenstein}},
  \bibnamefont{and} \bibinfo{author}{\bibfnamefont{E.~S.}
  \bibnamefont{Polzik}}, \bibinfo{journal}{Phys. Rev. Lett.}
  \textbf{\bibinfo{volume}{98}}, \bibinfo{pages}{100404}
  (\bibinfo{year}{2007}).

\bibitem[{\citenamefont{Eckert et~al.}(2008)\citenamefont{Eckert, Romero-Isart,
  Rodriguez, Lewenstein, Polzik, and Sanpera}}]{eckert2008}
\bibinfo{author}{\bibfnamefont{K.}~\bibnamefont{Eckert}},
  \bibinfo{author}{\bibfnamefont{O.}~\bibnamefont{Romero-Isart}},
  \bibinfo{author}{\bibfnamefont{M.}~\bibnamefont{Rodriguez}},
  \bibinfo{author}{\bibfnamefont{M.}~\bibnamefont{Lewenstein}},
  \bibinfo{author}{\bibfnamefont{E.~S.} \bibnamefont{Polzik}},
  \bibnamefont{and} \bibinfo{author}{\bibfnamefont{A.}~\bibnamefont{Sanpera}},
  \bibinfo{journal}{Nature Phys.} \textbf{\bibinfo{volume}{4}},
  \bibinfo{pages}{50} (\bibinfo{year}{2008}).

\bibitem[{\citenamefont{de~Vega et~al.}(2008)\citenamefont{de~Vega, Cirac, and
  Porras}}]{vega2008}
\bibinfo{author}{\bibfnamefont{I.}~\bibnamefont{de~Vega}},
  \bibinfo{author}{\bibfnamefont{J.~I.} \bibnamefont{Cirac}}, \bibnamefont{and}
  \bibinfo{author}{\bibfnamefont{D.}~\bibnamefont{Porras}},
  \bibinfo{journal}{Phys. Rev. A} \textbf{\bibinfo{volume}{77}},
  \bibinfo{pages}{051804} (\bibinfo{year}{2008}).

\bibitem[{\citenamefont{Douglas and Burnett}(2010)}]{douglas2010}
\bibinfo{author}{\bibfnamefont{J.~S.} \bibnamefont{Douglas}} \bibnamefont{and}
  \bibinfo{author}{\bibfnamefont{K.}~\bibnamefont{Burnett}},
  \bibinfo{journal}{Phys. Rev. A} \textbf{\bibinfo{volume}{82}},
  \bibinfo{pages}{033434} (\bibinfo{year}{2010}).

\bibitem[{\citenamefont{Mekhov et~al.}(2007)\citenamefont{Mekhov, Maschler, and
  Ritsch}}]{mekhov2007}
\bibinfo{author}{\bibfnamefont{I.~B.} \bibnamefont{Mekhov}},
  \bibinfo{author}{\bibfnamefont{C.}~\bibnamefont{Maschler}}, \bibnamefont{and}
  \bibinfo{author}{\bibfnamefont{H.}~\bibnamefont{Ritsch}},
  \bibinfo{journal}{Nature Phys.} \textbf{\bibinfo{volume}{3}},
  \bibinfo{pages}{319} (\bibinfo{year}{2007}).

\bibitem[{\citenamefont{Mekhov and Ritsch}(2009)}]{mekhov2009}
\bibinfo{author}{\bibfnamefont{I.~B.} \bibnamefont{Mekhov}} \bibnamefont{and}
  \bibinfo{author}{\bibfnamefont{H.}~\bibnamefont{Ritsch}},
  \bibinfo{journal}{Phys. Rev. Lett.} \textbf{\bibinfo{volume}{102}},
  \bibinfo{pages}{020403} (\bibinfo{year}{2009}).

\bibitem[{\citenamefont{Pino et~al.}(2011)\citenamefont{Pino, Wild, Makotyn,
  Jin, and Cornell}}]{Pino2011}
\bibinfo{author}{\bibfnamefont{J.~M.} \bibnamefont{Pino}},
  \bibinfo{author}{\bibfnamefont{R.~J.} \bibnamefont{Wild}},
  \bibinfo{author}{\bibfnamefont{P.}~\bibnamefont{Makotyn}},
  \bibinfo{author}{\bibfnamefont{D.~S.} \bibnamefont{Jin}}, \bibnamefont{and}
  \bibinfo{author}{\bibfnamefont{E.~A.} \bibnamefont{Cornell}},
  \bibinfo{journal}{Phys. Rev. A} \textbf{\bibinfo{volume}{83}},
  \bibinfo{pages}{033615} (\bibinfo{year}{2011}).

\bibitem[{\citenamefont{Slama et~al.}(2007)\citenamefont{Slama, Bux, Krenz,
  Zimmermann, and Courteille}}]{Slama2007}
\bibinfo{author}{\bibfnamefont{S.}~\bibnamefont{Slama}},
  \bibinfo{author}{\bibfnamefont{S.}~\bibnamefont{Bux}},
  \bibinfo{author}{\bibfnamefont{G.}~\bibnamefont{Krenz}},
  \bibinfo{author}{\bibfnamefont{C.}~\bibnamefont{Zimmermann}},
  \bibnamefont{and} \bibinfo{author}{\bibfnamefont{P.~W.}
  \bibnamefont{Courteille}}, \bibinfo{journal}{Phys. Rev. Lett.}
  \textbf{\bibinfo{volume}{98}}, \bibinfo{pages}{053603}
  (\bibinfo{year}{2007}).

\bibitem[{\citenamefont{{Brennecke} et~al.}(2007)\citenamefont{{Brennecke},
  {Donner}, {Ritter}, {Bourdel}, {K{\"o}hl}, and {Esslinger}}}]{Brennecke2007}
\bibinfo{author}{\bibfnamefont{F.}~\bibnamefont{{Brennecke}}},
  \bibinfo{author}{\bibfnamefont{T.}~\bibnamefont{{Donner}}},
  \bibinfo{author}{\bibfnamefont{S.}~\bibnamefont{{Ritter}}},
  \bibinfo{author}{\bibfnamefont{T.}~\bibnamefont{{Bourdel}}},
  \bibinfo{author}{\bibfnamefont{M.}~\bibnamefont{{K{\"o}hl}}},
  \bibnamefont{and}
  \bibinfo{author}{\bibfnamefont{T.}~\bibnamefont{{Esslinger}}},
  \bibinfo{journal}{Nature} \textbf{\bibinfo{volume}{450}},
  \bibinfo{pages}{268} (\bibinfo{year}{2007}).

\bibitem[{\citenamefont{{Colombe} et~al.}(2007)\citenamefont{{Colombe},
  {Steinmetz}, {Dubois}, {Linke}, {Hunger}, and {Reichel}}}]{Colombe2007}
\bibinfo{author}{\bibfnamefont{Y.}~\bibnamefont{{Colombe}}},
  \bibinfo{author}{\bibfnamefont{T.}~\bibnamefont{{Steinmetz}}},
  \bibinfo{author}{\bibfnamefont{G.}~\bibnamefont{{Dubois}}},
  \bibinfo{author}{\bibfnamefont{F.}~\bibnamefont{{Linke}}},
  \bibinfo{author}{\bibfnamefont{D.}~\bibnamefont{{Hunger}}}, \bibnamefont{and}
  \bibinfo{author}{\bibfnamefont{J.}~\bibnamefont{{Reichel}}},
  \bibinfo{journal}{Nature} \textbf{\bibinfo{volume}{450}},
  \bibinfo{pages}{272} (\bibinfo{year}{2007}).

\bibitem[{\citenamefont{Murch et~al.}(2008)\citenamefont{Murch, Moore, Gupta,
  and Stamper-Kurn}}]{Murch2008}
\bibinfo{author}{\bibfnamefont{K.~W.} \bibnamefont{Murch}},
  \bibinfo{author}{\bibfnamefont{K.~L.} \bibnamefont{Moore}},
  \bibinfo{author}{\bibfnamefont{S.}~\bibnamefont{Gupta}}, \bibnamefont{and}
  \bibinfo{author}{\bibfnamefont{D.~M.} \bibnamefont{Stamper-Kurn}},
  \bibinfo{journal}{Nature Phys.} \textbf{\bibinfo{volume}{4}},
  \bibinfo{pages}{561} (\bibinfo{year}{2008}).
 
 
 
 \bibitem{mattinson01}
 F. Mattinson, M. Kira, and S. Stenholm, J.\ Mod.\ Opt. {\bf 48}, 889 (2001).
 
\bibitem{larson05}
  J. Larson and E. Andersson,
  Phys. Rev. A {\bf 71}, 053814 (2005).

 

\bibitem[{\citenamefont{Walls and Milburn}(1994)}]{Walls1994}
\bibinfo{author}{\bibfnamefont{D.~F.} \bibnamefont{Walls}} \bibnamefont{and}
  \bibinfo{author}{\bibfnamefont{G.~J.} \bibnamefont{Milburn}},
  \emph{\bibinfo{title}{Quantum Optics}} (\bibinfo{publisher}{Springer},
  \bibinfo{address}{Berlin}, \bibinfo{year}{1994}).

\bibitem[{\citenamefont{Griessner et~al.}(2004)\citenamefont{Griessner, Jaksch,
  and Zoller}}]{Griessner2004}
\bibinfo{author}{\bibfnamefont{A.}~\bibnamefont{Griessner}},
  \bibinfo{author}{\bibfnamefont{D.}~\bibnamefont{Jaksch}}, \bibnamefont{and}
  \bibinfo{author}{\bibfnamefont{P.}~\bibnamefont{Zoller}},
  \bibinfo{journal}{J. Phys. B} \textbf{\bibinfo{volume}{37}},
  \bibinfo{pages}{1419} (\bibinfo{year}{2004}).

\bibitem[{\citenamefont{Van~Hove}(1954)}]{VanHove1954}
\bibinfo{author}{\bibfnamefont{L.}~\bibnamefont{Van~Hove}},
  \bibinfo{journal}{Phys. Rev.} \textbf{\bibinfo{volume}{95}},
  \bibinfo{pages}{249} (\bibinfo{year}{1954}).

\bibitem{Hansen2001}
  H. Hansen, T. Aichele, C. Hettich, P. Lodahl, A. I. Lvovsky, J. Mlynek, and S. Schiller, Opt. Lett. {\bf 26}, 1714 (2001).
  
\end{thebibliography}


\end{document}